\documentclass[%
 reprint,
superscriptaddress,
 amsmath,amssymb,
 aps,
 prl,
]{revtex4-1}

\usepackage{
	graphicx,
	amsmath,	
	amssymb,	
	amsthm,			
	fancyhdr,		
	gensymb,		
	sidecap,			
	subcaption,
	tikz,				
	listings,			
	color,			
	calc,				
	mdwlist,			
	thmtools,		
	etoolbox,		
	xspace,		
	bm,
	setspace,
	siunitx,
	dcolumn
}	
\newcommand{\iu}{{i\mkern1mu}}

\begin{document}

\title{Poynting singularities in the transverse flow-field of random vector waves}% Force line breaks with \\
\author{M.A. van Gogh}
\affiliation{Kavli Institute of Nanoscience, Delft University of Technology, 2600 GA Delft, The Netherlands}
\author{T. Bauer}
\affiliation{Kavli Institute of Nanoscience, Delft University of Technology, 2600 GA Delft, The Netherlands}
\author{L. De Angelis}
\affiliation{Kavli Institute of Nanoscience, Delft University of Technology, 2600 GA Delft, The Netherlands}
\affiliation{The Netherlands Institute for Neuroscience, Institute of the Royal Netherlands Academy of Arts and Sciences (KNAW), Amsterdam, The Netherlands}
\author{L. Kuipers}
\email{Corresponding author: l.kuipers@tudelft.nl}
\affiliation{Kavli Institute of Nanoscience, Delft University of Technology, 2600 GA Delft, The Netherlands}

\date{\today}

\begin{abstract}
In order to utilize the full potential of tailored flows of electromagnetic energy at the nanoscale, we need to understand its general behaviour given by its generic representation of interfering random waves. For applications in on-chip photonics as well as particle trapping, it is important to discern the topological features in the flow field between the commonly investigated cases of fully vectorial light fields and their 2D equivalents. We demonstrate the distinct difference between these cases in both the allowed topology of the flow-field and the spatial distribution of its singularities, given by their pair correlation function $g(r)$. Specifically, we show that a random field confined to a 2D plane has a divergence-free flow-field and exhibits a liquid-like correlation, whereas its freely propagating counterpart has no clear correlation and features a transverse flow-field with the full range of possible 2D topologies around its singularities.
\end{abstract}

\maketitle

\section{Introduction}

Historically, optical light fields were predominantly studied in the paraxial approximation or in propagating, fully vectorial structured beams  \cite{Born1999,Rubinsztein-Dunlop2017-short}. Future photonic applications, however, require knowledge about their on-chip behaviour \cite{Qiu2017-short,Feldmann2019-short,Kittlaus2018-short}, where the field is confined to two dimensions (2D). Understanding the structure of a light field and its flow of energy in these photonic systems is important for furthering our insight into the resulting intriguing effects \cite{RogueWaves-short}. One striking property of structured light fields in general is that they can exhibit a variety of optical singularities \cite{Gbur2016}. Singularities, in this context, are exceptional points in the describing field where a parameter becomes undefined, leading to a distinct topology of the field surrounding the singularity. Such singularities are ubiquitous in nature, occurring in a wide variety of systems, from black holes \cite{BH} to the bottom of a swimming pool \cite{Caustics}. 
In light, they can appear is as phase defects in a field component \cite{BerryNye74}, where the associated field amplitude vanishes. Such phase singularities are topological in nature and generically have an associated topological charge of $\pm 1$. Overall, the topological charge in a field has to be conserved \cite{Freund1998}, and as a result singularities can only be created or annihilated in pairs. The resulting network of singularities consequently makes up a topological skeleton of the field, from which the general behaviour of the entire field can be inferred \cite{Freund1993}.

It was recently shown that confinement of light to 2D leads to a fundamental difference in the behaviour of its 
singularities from those present in the transverse field of a paraxial beam \cite{DeAngelis2016,DeAngelis2018}. It can be presumed, that such a confinement not only has consequences for singularities in the complex vectorial electromagnetic fields, but also for the singular behaviour exhibited by other physical observables, such as the energy flow. This flow is commonly described by the Poynting vector \cite{OpticalCurrents} and is known to contain vortices, for example in the vicinity of the focus of a light beam \cite{Boivin1967} or near sub-wavelength apertures \cite{Schouten2003-short}.

In this letter, we reveal the topological changes of the Poynting vector field and its associated singularities when restricting the light field to 2D, in both experiment and simulations. While these singularities have already been experimentally observed in a chaotic billiard of microwaves \cite{Homann2009-short}, we here show their existence in a random wave ensemble in the optical regime, demonstrating how their behaviour changes with dimensionality. We highlight that the 2D confinement causes the in-plane flow-field to be divergence-free, thus limiting the allowed topological structures \cite{Dennis2001}. This constraint strongly influences the correlations between the singularities and leads to an intriguing cross-correlation between different singularity types.

\section{Generic topology of the flow field}
\label{sec:behaviour}

In complex scalar fields, phase singularities occur at positions where both the real and imaginary parts of the field become zero and the phase of the field increases in integers of $2 \pi$ around it and thus, their physical meaning is clear. Such an interpretation is less obvious in the flow-field, as it arises from a combination of electric and magnetic fields, resulting in a real-valued vector field given by

\begin{equation}\label{eqn:poynting}
\bm{S}= \frac{1}{2} \Re\left(\bm{E}\times\bm{H}^*\right).
\end{equation}
Here $\bm{E}$ denotes the electric field, $\bm{H}$ the magnetic field, with the star denoting complex conjugation. From Eqn. \ref{eqn:poynting} it follows that there are two clear cases when the magnitude of the Poynting vector vanishes: either all electric field components become zero, which we refer to as electric type singularities, or all magnetic field components become zero, which we call magnetic types. In the case of three-dimensional random waves, without additional symmetry, these cases do not occur generically, since it requires six components (3 real and 3 imaginary parts) to be simultaneously zero, with only five free parameters (three amplitudes and two relative phases) \cite{Berry2001}.
Additionally, the Poynting vector vanishes when $\bm{E}\times\bm{H}^*=0$ because the two fields are parallel to each other, or when $\bm{E}\times\bm{H}^*$ becomes purely imaginary. The first of these situations will again not occur in generic field distributions. In contrast, the case where the real part of the cross product vanishes is a generic feature of the flow field. We refer to singularities where the cross product vanishes, without either vector being fully zero, as polarization type singularities. For a full theoretical treatment of the possibilities for singularities in the three-dimensional Poynting field, see \cite{Novitsky}.

\begin{figure}[tbp]
\centering
{\includegraphics[width=\linewidth]{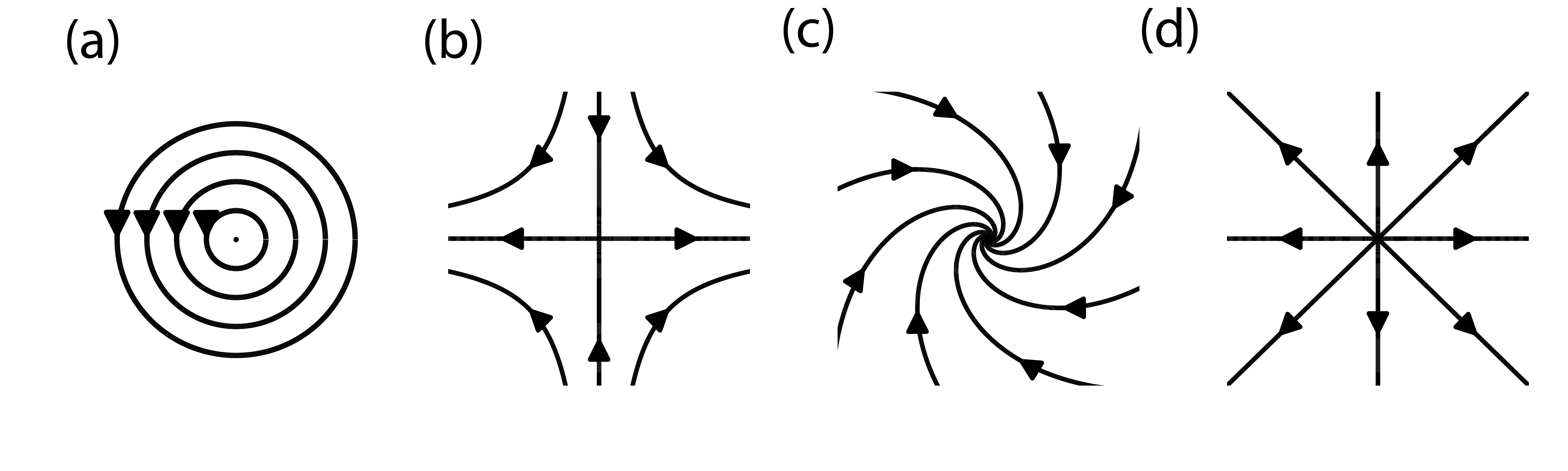}}
\caption{Critical points of a two-dimensional field. (a): Center. (b): Saddle point. (c): Focus. (d): Node. (See \cite{BEKSHAEV2007332})}
\label{fig:types}
\end{figure}

In order to visualise the flow field and its underlying topological structure, it is common to consider a projection of the full field onto a plane, resulting in the so-called transverse flow-field \cite{Gbur2016}. Since the transverse flow-field is a 2D real vector field, its singularities are the same as critical points, known from vector field topology \cite{Bifurcations}. Hence their possible local flow-field structures can be described by 2D vector field topologies, with Fig. \ref{fig:types} depicting the four possibilities for a generic topological structure around a critical point.

As a means to find locations of singularities in the transverse flow-field, we can artificially construct a complex scalar field out of the two real components of the flow-field in the plane as
\begin{equation}
    \xi=S_x+\iu S_y,
\end{equation}
analogous to the construction of a complex Stokes field to locate polarization singularities \cite{Freund2001, Dennis2002}. With this construction, we can analyze the singularities of the flow-field with the usual tools for phase singularities in a complex scalar field. The mapping $(S_x,S_y) \mapsto S_x+\imath S_y$ is a homeomorphism between $\mathbb{R}^2$ and $\mathbb{C}$, meaning that the topological properties of the constructed field are identical to that of the original vector field.

\section{Restriction to a 2D light field}

We investigate what happens to the flow-field singularities and their associated topologies when we restrict the field itself to 2D. We consider a single transverse electric (TE) mode confined to a plane, with a discrete in-plane wavenumber $k_\parallel$. Assuming the $x-y$ plane as the plane of propagation, the Poynting vector for a TE mode is given as
\begin{equation}\label{eqn:TEPoynting}
    \bm{S}_{\parallel,TE}=\Re\left[H_z^*(E_y,-E_x)\right].
\end{equation}
For this transverse field to be zero, the obvious cases are that either $H_z=0$, or $E_x=E_y=0$, both of which can occur in a generic TE field \cite{DeAngelis2016,DeAngelis2019}. Another option occurs when $H_z(E_y,-E_x)$ becomes completely imaginary, and hence the real part vanishes. This happens at points where the electric field is linearly polarized and the magnetic field is exactly $\pi/2$ out of phase with the electric field. These three cases are the same singularity types (magnetic, electric and polarization, respectively) introduced earlier, but restricted to the plane. Note that the magnetic type singularities are also phase singularities of $H_z$. Since the electric types require two field components to be zero, as opposed to only one for the magnetic types, they occur only very rarely ($< 1 \%$ of all singularities) \cite{DeAngelis2019}. Therefore we will not treat them further, although their occurrence is interesting on its own.

\begin{figure*}[t]
\centering
{\includegraphics[width=\linewidth]{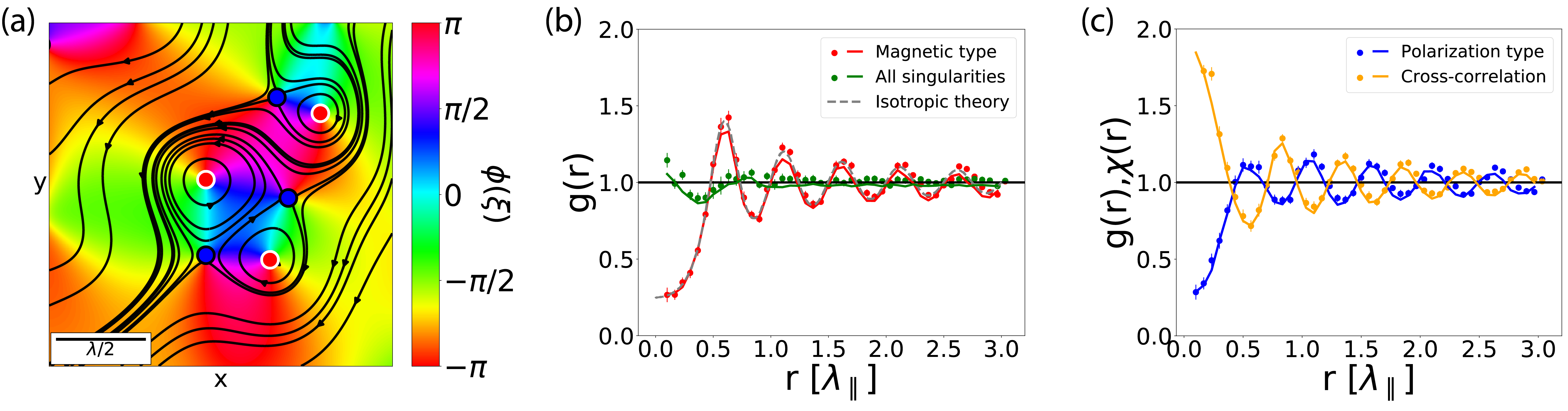}}
\caption{Topological structure and its associated pair correlations for a field confined to the plane. (a) shows the simulated topological structure, with the phase of the constructed complex scalar field $\xi$ shown in false colour. Red (Blue) dots are magnetic (polarization) type singularities. A white (black) outline of the dots are positive (negative) charged singularities. (b) The spatial distribution of magnetic type singularities (red) and the overall correlations between all singularities (green). Points are experimental data, and the solid lines are from numerical calculations. The gray dashed curve depicts the theoretical distribution for scalar isotropic random waves \cite{BerryDennis2000}. Distance is given in units of effective in-plane wavelength $\lambda_\parallel$. (c) The pair correlation $g(r)$ of the polarization types (blue) and the cross-correlation $\chi(r)$ between the magnetic and polarization types (orange).}
\label{fig:ringres}
\end{figure*}

In addition to the differentiation in singularity types, there can be restrictions on the allowed structure of the field surrounding them. Using Eqn. \ref{eqn:poynting} and Maxwell's equations in the absence of sources and currents, the divergence of the full 3D flow-field can be expressed as
\begin{equation}\label{eqn:divergence}
\nabla\cdot \bm{S}= \frac{1}{2}\Re\left[\iu \omega \left(\mu_0 |\bm{H}|^2 - \epsilon_0 |\bm{E}|^2\right) \right]=0.
\end{equation}  
This means that the divergence of the flow-field has to be identically zero everywhere in the field. Physically, the divergencelessness of the Poynting vector can be understood from the field consisting of a superposition of plane waves, each of which have a constant, divergence-free Poynting vector. Thus a linear sum of plane waves must also be divergence-free. Furthermore, the divergencelessness is just an expression of energy conservation, as points with a non-zero divergence would correspond to points where energy would enter or leave the system.
For a 2D confined field, the field is evanescently decaying in the out-of-plane direction ($S_z = 0$), meaning that $\partial_z S_z=0$, and thus:
\begin{equation}
    \nabla_\parallel\cdot \bm{S}_\parallel = 0.
\end{equation}
As a result, the only possible topologies are centers and saddles \cite{Dennis2001}, since nodes and spirals have a non-zero divergence. Additionally, it is easily verified that the rotation of the flow-field is not identically zero everywhere in the field. Hence the occurrence of centers is not forbidden. Indeed, from a topological point of view they are necessary: singularities have to be created and annihilated in pairs, which topologically manifests as a saddle-node bifurcation, where a saddle and a node are annihilated or created together \cite{Bifurcations}. In the case of a divergence-free field, this turns into a center-saddle bifurcation \cite{Broer2019}: creation/annihilation events cannot take place with just saddle points, and the field can either have no singularities, or must exhibit points with a different topology alongside the saddle points.

To experimentally study the occurrence and behaviour of the different topological structures in a generic 2D flow-field, we map the near fields of random TE waves propagating in a planar chaotic cavity \cite{DeAngelis2016} with a home-built polarization- and phase-resolving near-field optical microscope \cite{Burresi2009-short}. The employed interferometric detection approach \cite{Balistreri2000} allows us to map the in-plane TE-field components by raster scanning over a central $\SI{17}{\micro m} \times \SI{17}{\micro m}$ area of the chaotic cavity with a lateral step-size of approximately \SI{17}{nm}. From the in-plane distributions of $E_x$ and $E_y$, the $H_z$ component of the TE field can be reconstructed via Maxwell's equations, yielding the Poynting vector via Eqn. \ref{eqn:TEPoynting}. In order to compare experimental results with numerical calculations, we make use of random wave simulations. We use a set of 1000 plane waves with a random initial phase and direction to simulate a random TE wave field.

Fig. \ref{fig:ringres} (a) depicts a false-colour map for the phase of the constructed complex scalar field $\xi$ and with streamlines of the Poynting vector depicted in black for part of the simulated flow-field. This part is representative for the structures found in the field. We observe that for the topology of the singularities, which are marked with dots, only centers and saddle points are found in the transverse flow-field confined to the plane. 

To gain insight into the spatial distribution of the singularities that underpin the general structure of the flow-field, we determine their pair correlation, which tells us how the points are distributed spatially with respect to each other. For isotropic random waves, it is known that this function is liquid-like \cite{BerryDennis2000}, which has been verified experimentally \cite{DeAngelis2016,Stockmann,Homann2009-short}. Fig. \ref{fig:ringres} (b) shows the pair correlations for all singularities (depicted in green) and for only the magnetic type singularities with themselves (depicted in red). We find extremely good agreement between simulations (solid lines) and experiment (dots). The magnetic types exhibit a clear liquid-like correlation. Fig. \ref{fig:ringres} (b) also shows the theoretically computed distribution for isotropic scalar random waves, which is depicted by a dashed gray curve. Both the simulated and experimental results for the magnetic type are in excellent agreement with this curve. This behaviour is expected, as in a pure TE mode, $H_z$ should be isotropic in k-space \cite{DeAngelis2016}. On the other hand, when taking all singularities into account, the correlation between them is much weaker. The correlation only deviates slightly from unity at very short distances, after which it rapidly approaches unity.

In Fig. \ref{fig:ringres} (c) we present the pair correlation of only polarization types with themselves (depicted in blue) and the cross-correlations of the magnetic and polarization types (depicted in orange), where we look at the pair correlations between two singularities of opposite type. The polarization type singularities, like the magnetic type, exhibit a liquid-like correlation, but with a smaller first-peak amplitude. This dampening behaviour was already observed from the approximate model for the pair correlations of random waves \cite{Homann2009-short}. Such striking difference with the isotropic wave model hints at a deeper physical distinction between the singularity types. The cross-correlation, in contrast, shows a strong anti-correlation. Distances where there is a higher probability to find a singularity of the same type have a lower probability to find one of the opposite type and vice-versa. This explains why each type exhibits a clear correlation amongst themselves, while the overall distribution does not.

In the discussion above we considered a confined TE mode, which has a transverse wavenumber $k_\parallel$ outside the light cone, meaning that $k_\parallel > k_0$, where $k_0^2={k^2_\parallel+k^2_\perp}$.  However, the results stay unchanged if $k_\parallel < k_0$ as long as a single transverse wavenumber is considered. Hence the results not only hold for a confined field, but for any field that is invariant along the z-direction, such as a Bessel beam.

\section{Fields with varying in-plane wavenumber}

Finally, we investigate what happens when we ease the restriction of a discrete in-plane wavenumber. By allowing for a range of in-plane wavenumbers to build up the field, $k_z$ also has to vary. Therefore, the field is in general no longer invariant along the z-axis, meaning the out-of-plane divergence is not automatically equal zero. Since the full flow-field has to be divergence-free, this necessitates a non-zero in-plane divergence. Thus, the transverse flow-field exhibits singularities that may have a topology differing from centers and saddles.

\begin{figure}[t]
\centering
{\includegraphics[width=\linewidth]{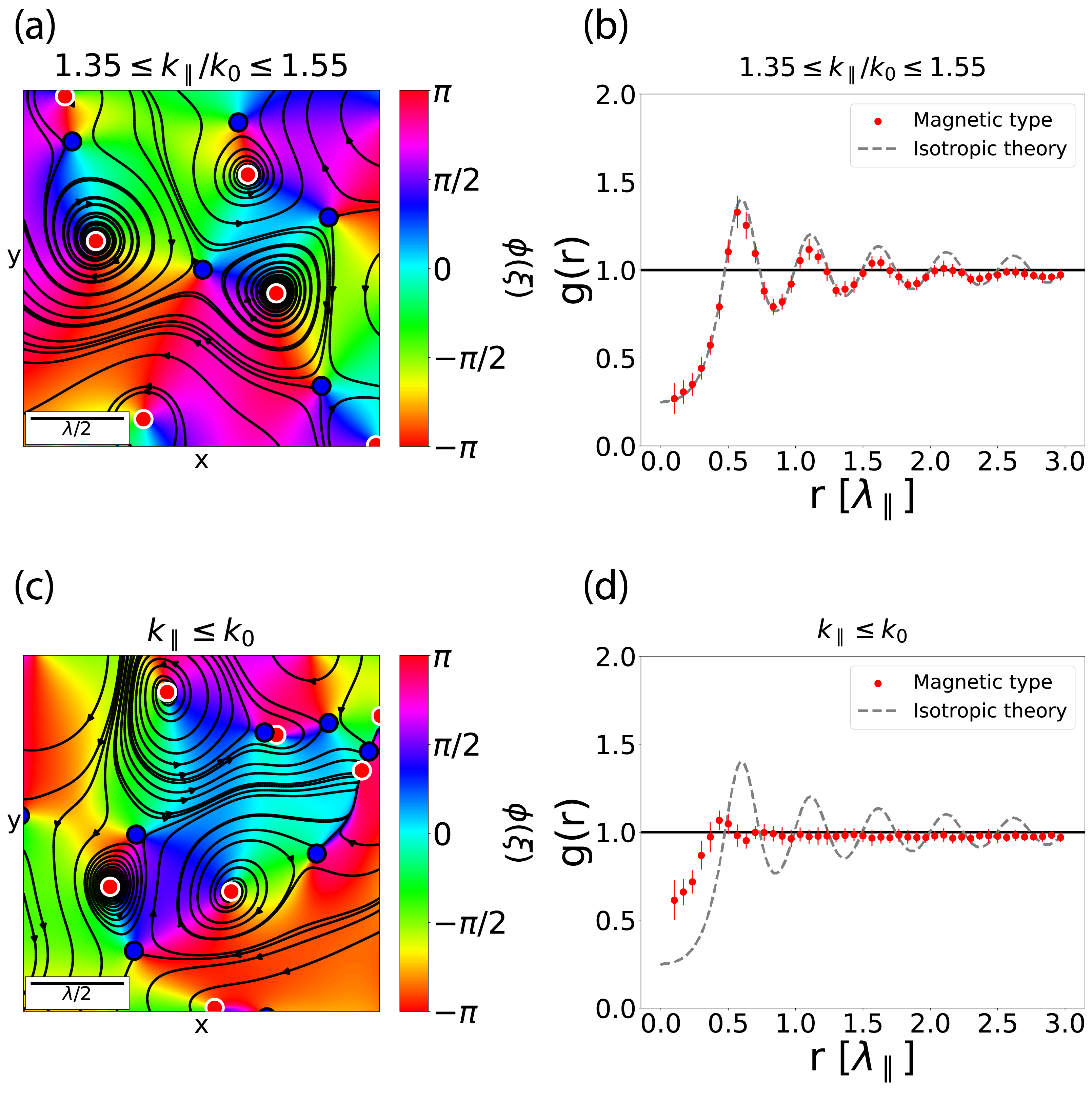}}
\caption{Results for fields with varying in-plane wavenumber, which can either be inside ($k \leq k_0$), or outside ($k > k_0$) the light cone. (a) and (c) show the topological behaviour, with the phase of the scalar field shown in false colour. Red (Blue) dots are magnetic (polarization) type singularities. A white (black) outline of the dots are positive (negative) charged singularities.  (b) and (d) show the spatial distribution of magnetic type singularities. The gray curve shows the theoretical distribution for isotropic random waves. The red dots are simulated points. Distance is given in units of effective in-plane wavelength.}
\label{fig:diskres}
\end{figure}

Fig. \ref{fig:diskres} (a) and (c) depict simulations of the transverse flow-field for different in-plane wavenumber ranges, utilizing the same false colour map and streamlines as in Fig. \ref{fig:ringres}(a). The presented parts are representative for the global field. The difference in apparent singularity density is coincidental. Fig. \ref{fig:diskres} (b) and (d) show the pair correlation function for the magnetic type singularities (depicted in red), and the theory for isotropic scalar random waves (gray dashed curve).

Comparing Fig. \ref{fig:diskres} (b) and (d), we find significant differences in the correlations for fields that have a varying in-plane wavenumber. For in-plane wavenumbers of $1.35 \leq k_\parallel/k_0 \leq 1.55$ (Fig. \ref{fig:diskres} (b) ), the correlation for magnetic type singularities still exhibits a liquid-like behaviour. However, its oscillations around unity decay more rapidly, making for an effectively shorter correlation length given by the size of the chosen wavenumber spread. Fig. \ref{fig:diskres} (d) depicts the correlation for a freely propagating light field, i.e. with allowed in-plane wavenumbers of $0 \leq k_\parallel/k_0 \leq 1$. The correlation swiftly approaches unity after roughly half a wavelength distance, and does not exhibit any clear oscillations. Thus, we deduce that the magnetic type singularities only exhibit extremely short nearest neighbour correlations.

Therefore, when allowing the in-plane wavenumber to vary, we see a striking difference between the case of an imaginary or purely real out-of-plane wavenumber. When it is imaginary ($k_\parallel > k_0$), we are bound by the same topological restrictions as for the 2D case and the singularity correlations are still present, but with a shorter correlation length than for isotropic random waves. When the out-of-plane wavenumber is real ($k_\parallel \leq k_0$), there are no restrictions on the topology anymore and no clear correlations can be observed.

\section{Conclusions}
We have investigated the topological properties and spatial distributions of singularities in the transverse flow-field of random waves, using theory, simulations and experimental data. Specifically, we demonstrated experimentally that by confining the wave field to two dimensions, the topological structure of the flow-field is restricted to only saddle and center type singularities. This restriction holds for any z-invariant light field and manifests itself in a liquid-like pair correlation of the magnetic-type singularities of the flow-field, as well as a distinct cross-correlation between all singularity sub-species. When loosening the restriction of a single transverse wavenumber, a significant difference occurs between fields originating from wavevectors fully outside or inside the light cone. For propagating beams such as a paraxial or non-perfect Bessel beam, the full range of possible 2D topological behaviour is found. This finding was corroborated by the change in the pair correlation function of the two different cases. For a propagating field, no clear pair correlation is present. On the other hand, the pair correlation of any confined field is reminiscent of a liquid-like state, but with the amplitude of the oscillations around unity decaying faster than in the case of isotropic random waves. The discussed restrictions on the generically occurring topologies in the transverse flow-field of 2D confined light might lead to novel concepts for utilizing the in-plane optical momentum present in many photonic systems for enhanced particle manipulation schemes.

\section{Acknowledgements}
We thank Andrea Di Falco for fabricating the sample used in the near-field experiments.

\section{Funding Information}
The authors acknowledge funding from the European Research Council (ERC Advanced Grant No. 340438-CONSTANS).

\end{document}